\documentclass[9pt]{extarticle}
\usepackage{spconf,graphicx}
\usepackage{multirow}
\usepackage{booktabs}
\usepackage{ctable}
\usepackage{cite}
\usepackage{amsmath, amssymb, amsthm, enumitem}

\title{HiFi-SR: A Unified Generative Transformer-Convolutional Adversarial Network for High-Fidelity Speech Super-Resolution}
\name{%
\begin{tabular}{@{}c@{}}
Shengkui Zhao  \quad %$^{\star \dagger}$ \qquad 
Kun Zhou \quad
Zexu Pan \quad
Yukun Ma\quad %$^{\star}$ \qquad 
Chong Zhang  \quad %$^{\star \dagger}$ \qquad 
Bin Ma
\end{tabular}}
\address{Tongyi Lab, Alibaba Group, Singapore\\
	\{shengkui.zhao, b.ma\}@alibaba-inc.com\\}
\begin{document}
\setlength{\abovedisplayskip}{3pt}
\setlength{\belowdisplayskip}{3pt}
%\ninept
%
\maketitle
\begin{abstract}
The application of generative adversarial networks (GANs) has recently advanced speech super-resolution (SR) based on intermediate representations like mel-spectrograms. However, existing SR methods that typically rely on independently trained and concatenated networks may lead to inconsistent representations and poor speech quality, especially in out-of-domain scenarios. In this work, we propose HiFi-SR, a unified network that leverages end-to-end adversarial training to achieve high-fidelity speech super-resolution. Our model features a unified transformer-convolutional generator designed to seamlessly handle both the prediction of latent representations and their conversion into time-domain waveforms. The transformer network serves as a powerful encoder, converting low-resolution mel-spectrograms into latent space representations, while the convolutional network upscales these representations into high-resolution waveforms. To enhance high-frequency fidelity, we incorporate a multi-band, multi-scale time-frequency discriminator, along with a multi-scale mel-reconstruction loss in the adversarial training process. HiFi-SR is versatile, capable of upscaling any input speech signal between 4 kHz and 32 kHz to a 48 kHz sampling rate. Experimental results demonstrate that HiFi-SR significantly outperforms existing speech SR methods across both objective metrics and ABX preference tests, for both in-domain and out-of-domain scenarios.
\end{abstract}   
\section{Introduction}
\label{sec:intro}
Speech super-resolution (SR) aims to reconstruct a high-resolution speech signal from a low-resolution input that retains only a portion of the original samples. Also referred to as bandwidth extension, this process enriches low-frequency content with high-frequency details. High-resolution speech signals, such as those at 48 kHz, not only deliver a superior listening experience but also improve speech intelligibility. Consequently, SR is a crucial technique for enhancing the quality of low-resolution speech, with applications in speech quality enhancement \cite{Chennoukh2001}, historical recording restoration \cite{Liu2021V}, and text-to-speech synthesis \cite{Nakamura2014A}.

Speech SR is particularly challenging due to the need to manage the high temporal resolution of speech signals, which contain structural patterns across various time scales with both short- and long-term dependencies. Early research in this field primarily relied on statistical methods, leading to slow progress \cite{Cheng1994S, Pulakka2011S, Eldin2011M, Turan2015S}. Recently, learning-based approaches using deep neural networks (DNNs) have shown promising advancements. Most learning-based methods focused on non-generative networks with a target resolution of 16 kHz \cite{Kuleshov2017A, Lim2018T, Li2019S, Hou2020S, Wang2021T}. For example, AECNN \cite{Wang2021T} utilized an autoencoder for waveform-to-waveform mapping, while TFNet \cite{Lim2018T} employed dual-branch convolutional neural networks (CNNs) that perform mapping in both time and frequency domains. 
More recent studies have successfully adopted generative models to achieve higher target resolutions of 48 kHz \cite{Lee2021NU, Zhang2021W, Han2022NU, Liu2022N, liu2023a}. NU-WAV \cite{Lee2021NU} utilizes a diffusion probabilistic model to generate high-resolution waveforms from low-resolution inputs. WSRGlow \cite{Zhang2021W} employs a glow-based generative model to generate high-resolution samples conditioned on low-resolution inputs. While both NU-WAV and WSRGlow have succeeded in achieving 48 kHz super-resolution, they are constrained by their ability to train on only one fixed input sampling rate at a time. Additionally, their performance falls short compared to the latest models NVSR \cite{Liu2022N} and AudioSR \cite{liu2023a}, which leverage generative adversarial networks (GANs) and mel-spectrogram representation. Both NVSR and AudioSR decompose the task into two stages: predicting high-resolution mel-spectrograms from low-resolution ones and then reconstructing the time-domain waveform from the high-resolution mel-spectrogram. 
%The key difference between NVSR and AudioSR lies in the first stage where NVSR uses a convolutional ResUnet \cite{liu2021C}, whereas AudioSR employs a latent diffusion model (LDM). 
We find that dividing the SR task into separate steps can introduce inconsistent representations. For instance, the output mel-spectrogram from the first stage may not be optimally aligned with the vocoder’s input requirements, potentially affecting the output quality. Furthermore, when the input speech differs significantly from the training data, the separately trained models may struggle to generalize effectively.
%, As evidenced in our experiments, this issue is especially problematic in out-of-domain scenarios, where input speech differs significantly from the training data, as the separately trained models may struggle to generalize effectively. Furthermore, using separate training processes increases the demand for training resources.  
\begin{figure*}
  \centering
  \includegraphics[width=15cm]{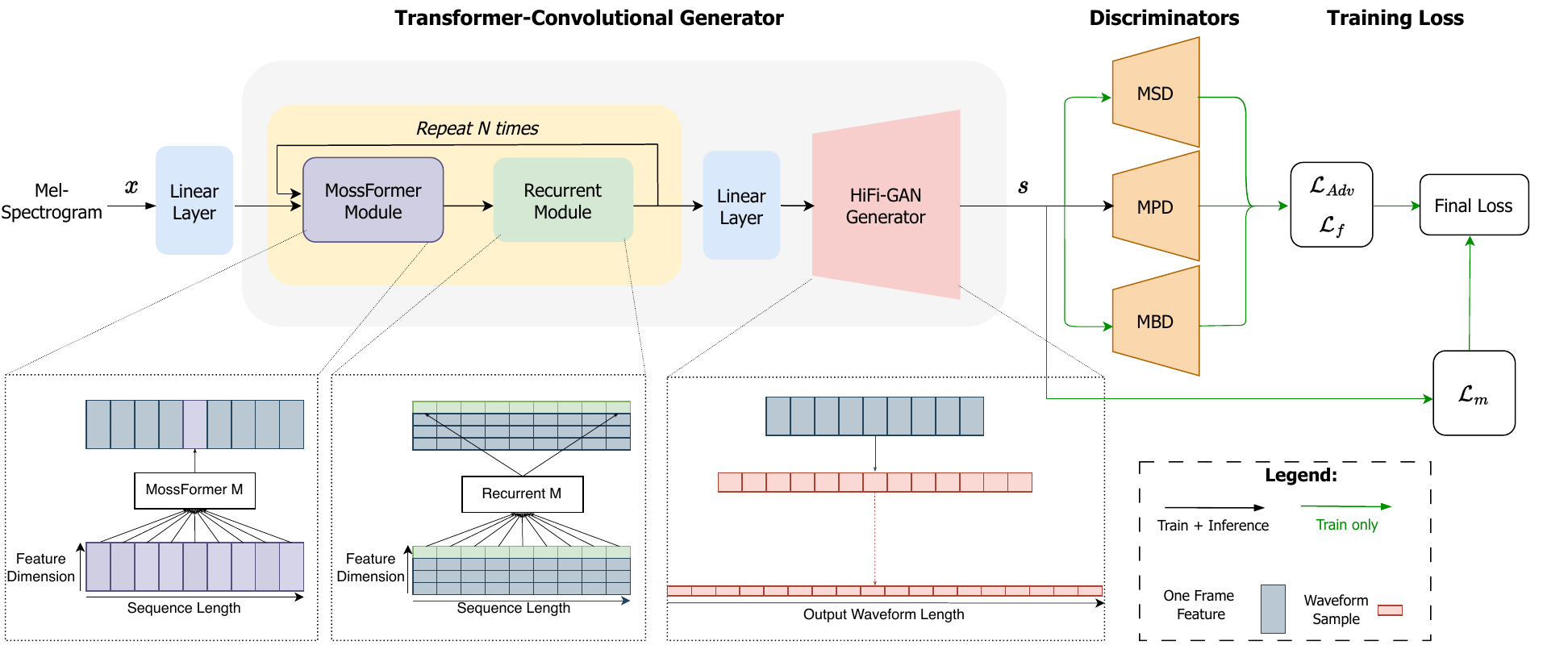}
%  \vspace{-3mm}
  \caption{Overview of our proposed generative transformer-convolutional adversarial network for speech super-resolution (HiFi-SR). The transformer-convolutional generator includes a hybrid MossFormer and recurrent network followed by a reused HiFi-GAN generator. Three discriminators of MSD, MPD and MBD are combined with feature matching loss $\mathcal{L}_f$ and mel-spectrogram loss $\mathcal{L}_m$ for high-fidelity adversarial training.}
  \label{fig1}
%\vspace{-3mm}
\end{figure*} 
%The transformer network consists of a hybrid MossFormer and recurrent module, repeated $N$ times. The MossFormer module processes the entire sequence to capture global dependencies, whereas the recurrent module focuses on learning temporal patterns across each embedding dimension. The HiFi-GAN generator network up-samples the latent representations to produce high-resolution waveforms.

\vspace{-0.5mm}
In this work, we propose a unified network that leverages end-to-end adversarial training to achieve high-fidelity and more generalized speech super-resolution at 48 kHz. Unlike NVSR and AudioSR, our approach features a unified transformer-convolutional generator that seamlessly handles both the prediction of latent representations and their conversion into time-domain waveforms. This design allows the latent representations to move beyond mel-spectrogram constraints, enabling the transformer network to optimize them for optimal alignment with the convolutional network during waveform generation. The transformer network taking from MossFormer2 \cite{zhao2023m2} is particularly effective at capturing long-term dependencies, beneficial for inferring high-frequency structures, making it a proper encoder choice for converting low-resolution mel-spectrograms into latent space representations. Our convolutional network, using the HiFi-GAN generator \cite{kong2020hifi}, ensures high-quality waveform generation. To further enhance high-frequency fidelity, we incorporate a multi-band, multi-scale time-frequency discriminator and a multi-scale mel-reconstruction loss within the adversarial training framework. We demonstrate that our proposed approach, termed HiFi-SR, can upscale any input speech signal between 4 kHz and 32 kHz to a 48 kHz sampling rate. Experimental results show that HiFi-SR significantly outperforms existing speech SR methods across both objective metrics and ABX preference tests, in both in-domain and out-of-domain scenarios.
\section{Method}
%When using a mel-spectrogram as input and generating waveform output, our proposed HiFi-SR model employs similar optimization strategies as neural vocoders like MelGAN \cite{kumar2019melgan} and HiFi-GAN. 
%One might consider directly applying neural vocoders for SR, as suggested in \cite{Liu2022N}. However, 
%These vocoders primarily focus on the mel-spectrogram inversion for waveform reconstruction.
%, with their experiments typically conducted on speech signals below 24 kHz. 
%In contrast, SR requires not only waveform reconstruction but also accurate high-resolution prediction. 
%Therefore, additional efforts are needed to enhance both the generator network and the training strategy to achieve high-fidelity SR.
When using a mel-spectrogram as input to generate waveform output, our proposed HiFi-SR model adopts optimization strategies similar to neural vocoders like MelGAN \cite{kumar2019melgan} and HiFi-GAN, which primarily focus on mel-spectrogram inversion for waveform reconstruction. However, SR requires not only waveform reconstruction but also precise high-resolution prediction.
\subsection{Transformer-Convolutional Generator}
To this end, we propose replacing the fully convolutional generators found in HiFi-GAN with a transformer-convolutional generator as shown in Figure 1. Our generator combines a transformer network and a convolutional feed-forward network, taking mel-spectrogram $s$ as input and producing raw waveform $x$ as output. To accommodate varying input sampling rates, we first up-sample signals with lower sampling rates to 48 kHz before extracting mel-spectrograms. 
%The input size for the generator is \textit{batch} $\times$ $80 \times L$, where $L$ is the temporal frame length.
%In this work, we use 80-band mel-spectrograms with a $256\times$ lower temporal resolution. Therefore, the input size for the generator is \textit{batch} $\times$ $80 \times L$, where $L$ is the temporal frame length. 
Our transformer network reuses the MossFormer2 block developed in our previous work \cite{zhao2023m2}. The MossFormer2 block is repeated $N$ times to enhance the modelling capability. Before the first block, the mel-spectrogram is projected into a higher-dimensional space using a linear layer. As detailed in \cite{zhao2023m2, zhao2023moss}, each MossFormer2 block combines a MossFormer and a recurrent block. The MossFormer component employs joint local and global self-attention to fully capture long-term global dependencies within the input sequence. It also utilizes an attentive gating mechanism that reduces the number of self-attention heads to one, significantly simplifying the multi-head attention requirement. The recurrent block, based on the feedforward sequential memory network (FSMN) \cite{Zhang2018D}, incorporates dilations to achieve broader receptive fields. This recurrent block is crucial for capturing recurrent patterns related to phonetic structures, prosody, and semantic associations in speech signals, thereby improving the prediction accuracy of high-frequency details. 

The transformer network outputs an enriched latent representation of the mel-spectrogram input, which is then fed into a convolutional network for waveform synthesis. Our convolutional network is based on the HiFi-GAN generator \cite{kong2020hifi}, consisting of a series of transposed convolutional layers that upsample the input sequence until the output sequence length matches that of the high-resolution waveform. Each transposed convolutional layer is followed by a multi-receptive field fusion (MRF) module. The MRF module is used to capture patterns of varying lengths by summing outputs from multiple residual blocks, each with different kernel sizes and dilation rates to create diverse receptive field patterns. We adjusted the hidden dimension $h_u$, transposed convolution kernel sizes $k_u$, MRF kernel sizes $k_r$, and MRF dilation rates $D_r$ for optimal performance in our SR experiments.
\begin{figure*}[t]
  \centering
  \includegraphics[width=17.5cm]{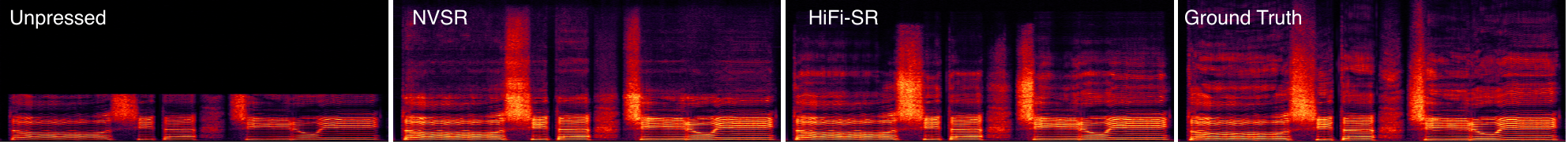}
  \vspace{-3mm}
  \caption{Spectrogram illustrations of different system outputs for a sample input from the VocalSet singing test set. It demonstrates that HiFi-SR significantly outperforms the baseline NVSR model.}
  \label{fig1}
\vspace{-3mm}
\end{figure*} 
\subsection{Discriminator Design}
As demonstrated in MelGAN and HiFi-GAN, the design of the discriminator is critical for generating high-fidelity audio waveforms. We utilize the multi-scale discriminator (MSD) from MelGAN and the multi-period discriminator (MPD) from HiFi-GAN to capture periodic speech patterns at different levels. The MSD operates on three input scales $1\times$, $2\times$, and $4\times$ using average pooling, while the MPD processes disjoint samples with periods of [2,3,5,7,11]. While both MSD and MPD contribute to high audio fidelity, we observe that over-smoothing artifacts can still appear in the high-frequency regions of the generated spectrograms. The multi-resolution discriminator (MRD) proposed in BigVGAN \cite{lee2023b} could mitigate such artifacts by operating on linear spectrograms. However, MRD discards phase information, limiting its ability to penalize phase modeling errors at high frequencies.

To address these issues, we adopt a multi-band, multi-scale time-frequency discriminator (MBD), inspired by audio codec discriminators \cite{Copet2022H, kumar2023hi}. The MBD takes the concatenated real and imaginary parts of the complex short-time Fourier transform (STFT) as input. We use five STFT window lengths [4096,2048,1024,512,256], with frequency bands split at [0.0,0.1,0.25,0.5,0.75,1.0]. Each time scale and sub-band shares identical network blocks, consisting of a 2D convolutional layer with a $3\times 8$ kernel and 32 channels, followed by 2D convolutions with dilation rates of $1$, $2$, and $4$ in the time dimension, and a stride of 2 along the frequency axis. A final 2D convolution with a $3\times 3$ kernel and a stride of $(1, 1)$ generates the final prediction. In our SR experiments, we combine MSD, MPD, and MBD for enhanced performance.
\subsection{Training Objective}
To optimize the generator and the discriminators, our training loss combines GAN loss, multi-scale mel-spectrogram loss, and feature matching loss, as detailed below. 

\hspace{-15pt}\textbf{GAN Loss:} For our generator and discriminators, we employ the least-squares objective from LS-GAN \cite{Mao2017LS}, which has proven highly effective for adversarial training. The losses for MSD, MPD, and MBD are computed in the same manner, with their individual losses summed to form the final discriminator loss: 
\begin{align}
\mathcal{L}_{Adv}(D)&=\sum_{i=1}^3\sum_{k=1}^{K_i}{\mathbb{E}_{(x,s)}\Big[(1-D_{i,k}(x))^2+(D_{i,k}(G(s)))^2\Big]}, \\
\mathcal{L}_{Adv}(G)&=\sum_{i=1}^3\sum_{k=1}^{K_i}\mathbb{E}_s\Big[(1-D_{i,k}(G(s)))^2\Big].
\end{align}
Here, $D_{i,k}$ denotes a sub-discriminator, where $i=1,2,3$ corresponds to the three discriminator types of MSD, MPD, and MBD, and $k$ refers to the $k$-th scale or band. $K_i$ represents the total number of scales or bands for the $i$-th discriminator.

\hspace{-15pt}\textbf{Multi-Scale Mel-Spectrogram Loss:}  In addition to the GAN loss, we incorporate a multi-scale mel-spectrogram loss to promote frequency modeling across multiple time scales, as suggested for codecs \cite{kumar2023hi}. The mel-spectrogram loss is known to improve stability, fidelity, and convergence speed \cite{kong2020hifi}. In our model, we apply an L1 loss across 7 mel-spectrogram bins [5,10,20,40,80,160,320], computed using window lengths of [32,64,128,256,512,1024,2048] with a hop length of $w_j / 4$, where $\{w_j, j=1,2,...,7\}$ represents the different window lengths. The multi-scale mel-spectrogram loss is defined as:
\begin{equation}
\mathcal{L}_{m}(G)=\sum_{j=1}^7{\mathbb{E}_{(x,s)}\Big[\|\text{Mel}_j(x)-\text{Mel}_j(G(s))\|_1\Big]}
\end{equation} 

\hspace{-15pt}\textbf{Feature Matching Loss:} We also incorporate a feature matching loss to stabilize the training process. As demonstrated in \cite{kong2020hifi}, this loss improves the quality of generated outputs by ensuring that the generator produces feature representations similar to those of real data at various levels within the discriminators. The feature matching loss is defined as:
\begin{equation}
\mathcal{L}_f(G)=\sum_{i=1}^3\sum_{k=1}^{K_i}{\mathbb{E}_{(x,s)}\Big[\frac{1}{L_i}\sum_{l=1}^{L_i}\frac{1}{T_{i,k}^l}\|D_{i,k}^l(x)-D_{i,k}^l(G(s))\|_1\Big]},
\end{equation}
where $L_{i,k}$ denotes the number of layers in the $\{i,k\}$-th discriminator, $D_{i,k}^l$ and $T_{i,k}^l$ denote the output feature and the feature length in the $l$-th layer of the $\{i,k\}$-th discriminator.

\hspace{-15pt}\textbf{Final Loss:} The final objectives for the generator and discriminators are defined as follows:
\begin{align}
\mathcal{L}_G&=\mathcal{L}_{Adv}(G)+\lambda_m\mathcal{L}_{m}(G)+\lambda_f\mathcal{L}_f(G), \\
\mathcal{L}_D&=\mathcal{L}_{Adv}(D)
\end{align}
where we set $\lambda_m=7$ and $\lambda_f=1.5$ to balance the weighted losses. 
\section{Experiment}
\subsection{Dataset}
To evaluate our proposed approach, we created a training set from the VCTK speech corpus \cite{Yamagishi2019C}, which includes recordings from 108 English speakers with a total of 44 hours of speech at 48 kHz. Consistent with the data preparation strategy used in \cite{Liu2022N}, we used recordings from 100 speakers for training and the remaining 8 speakers for testing.  To assess the generalizability of HiFi-SR to unseen speech types and data types, we created two additional test sets. The EXPRESSO dataset \cite{nguyen2023exp}, containing 17 hours of expressive reading speech from 4 North American English speakers, was used, with 10\% of recordings from each speaker and style forming a 1.7-hour EXPRESSO test set. The VocalSet \cite{Wilkins2018VocalSetAS}, a dataset of a cappella singing voices from 20 professional singers (11 male, 9 female), was also used, with recordings from 2 male and 2 female singers making up a 2-hour VocalSet test set.
\begin{table}
\center
\footnotesize
\caption{Objective evaluation results for 48 kHz speech super-resolution from input sampling rates of 4 kHz, 8 kHz, 16 kHz, and 24 kHz on the VCTK test set. The evaluation metric is the average LSD across all utterances, with lower values indicating better performance. Nu-wave and WSRGlow have fixed input resolutions.}
%\vspace{-2.5mm}
\setlength\tabcolsep{4.0pt} % default value: 6pt
\begin{tabular}{lcccccc}
\specialrule{.1em}{.05em}{.05em}
{Model} &{No. Parameters} &{4 kHz} & {8 kHz} &{16 kHz} & {24 kHz} &{AVG}\\ \hline
%\multirow{2}{*}{Model} &\multirow{2}{*}{$\alpha$} &\multirow{2}{*}{$\beta$}& \multicolumn{3}{c}{Ob Metrics}         \\
%\cline{4-6}
Unprocessed	  &- &6.08	       &5.15        &4.85     &3.84      &4.98   \\ \hline
Nu-wave       &3.0M$\times$4 &1.42	           &1.42        &1.36      &1.22   &1.36    \\ 
WSRGlow       &229.9M$\times$4 &1.12			   &0.98		&0.85	   &0.79  &0.94      \\
AudioSR-Speech  &- &1.15 		    &1.03       &0.82      &0.69      &0.92  \\ 
NVSR          &99.0M &0.98	           &0.91		&0.81      &0.70     &0.85    \\ \hline
HiFi-SR w/o MBD                   &101M  &0.97	           &0.88		&0.79      &0.69   &0.83      \\
HiFi-SR w/o $\mathcal{L}_{m}(G)$  &101M  &0.98	           &0.89		&0.80      &0.70    &0.84     \\
HiFi-SR (proposed)                &101M  &\textbf{0.95} 	&\textbf{0.86}  &\textbf{0.77}  &\textbf{0.68} &\textbf{0.82} \\ 
\specialrule{.1em}{.05em}{.05em}
\end{tabular}
\vspace{-3mm}
\end{table}  
\begin{figure}
  \centering
  \includegraphics[width=8.5cm]{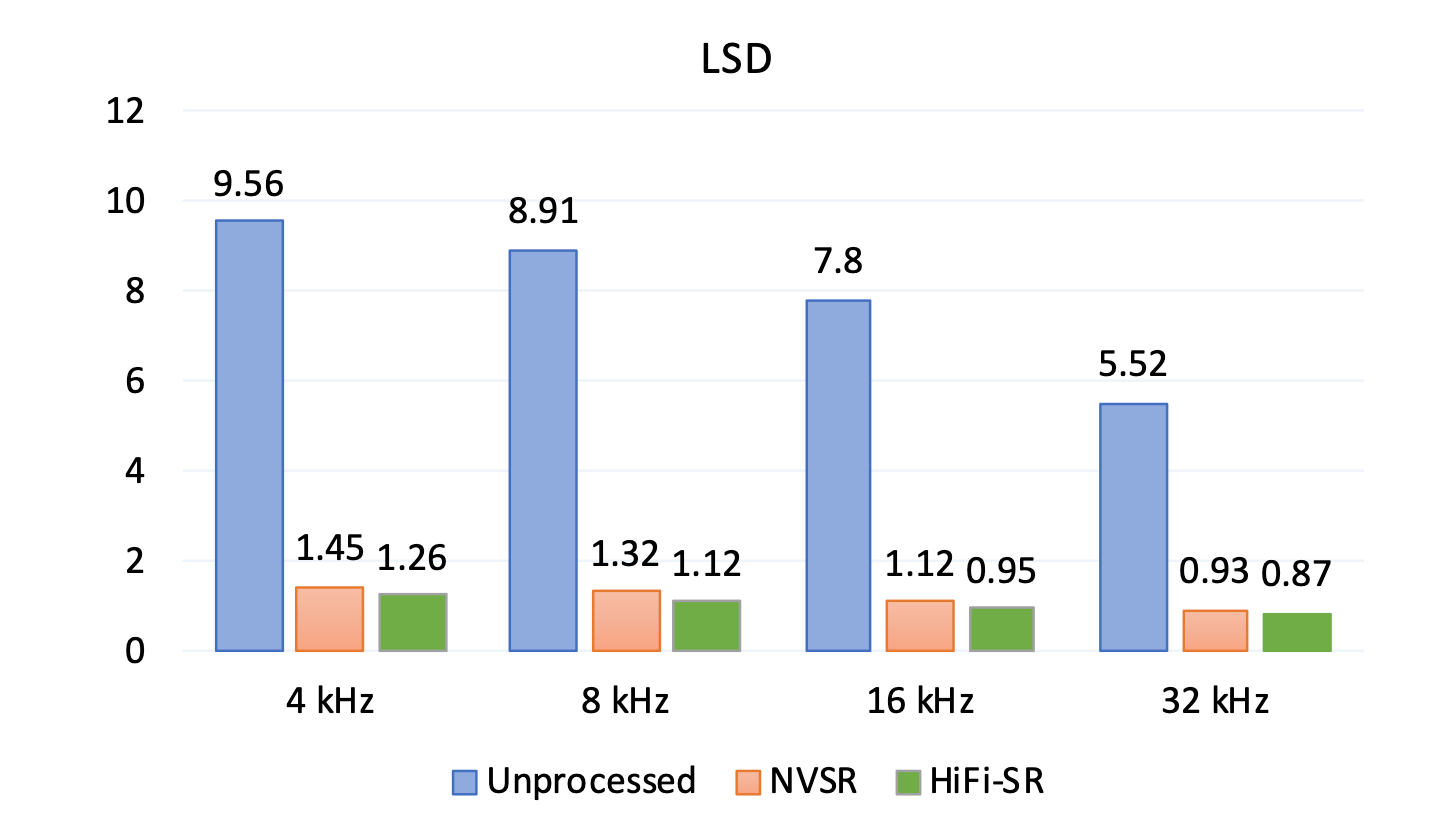}
  \vspace{-3mm}
  \caption{Comparison results of NVSR and HiFi-SR on EXPRESSO test set with 48 kHz target sampling rate and four input sampling rates.}
  \label{fig1}
\vspace{-3mm}
\end{figure} 
\subsection{Evaluation Metrics}
For the objective evaluation metric, we use Log-spectral distance (LSD) to evaluate the SR performance  following \cite{Liu2022N, liu2023a}.  Let $\mathbf{S}$ and $\hat{\mathbf{S}}$ stand for the magnitude spectrograms of the target speech $s$ and the generated speech $\hat{s}$. LSD is defined as follows:
\begin{equation}
\text{LSD}(\mathbf{S}, \hat{\mathbf{S}}) = \frac{1}{T}\sum_{t=1}^T\sqrt{\frac{1}{F}\sum_{f=1}^F \Big[\text{log}_{10}\frac{\mathbf{S}(t,f)^2}{\hat{\mathbf{S}}(t,f)^2}\Big]^2}
\end{equation}
%\Big(\frac{\mathbf{S}(t,f)^2}{\hat{\mathbf{S}}(t,f)^2}\Big)^2
LSD is a frequency-domain metric that measures the logarithmic distance between two magnitude spectra in dB. When the two spectra are identical, LSD reaches its minimum value of 0 dB. We report the average LSD across all tested audio files. For subjective evaluation, we conducted an ABX listening test, where raters selected their preferred audio output based on sound quality. Eight listeners participated in the test, each evaluating 50 audio pairs.
\subsection{Traning Details}
Our baseline models include Nu-wave, WSRGlow, NVSR, and AudioSR, all targeting a sampling rate of 48 kHz. For the VCTK test set, we used the baseline results reported in their respective publications. For the EXPRESSO and VocalSet test sets, we employed the NVSR pre-trained models based on the open-source code\footnote{https://github.com/haoheliu/ssr\_eval}. Following the method described in \cite{Liu2022N}, we simulated training and test sets by applying various low-pass filters to 48 kHz audio data to obtain lower sampling rates between 4 kHz and 32 kHz. We used 80-band mel-spectrograms with a $256\times$ lower temporal resolution. For the HiFi-SR model setup, we used $N=24$ MossFormer2 blocks with embedding size of 512. In the HiFi-GAN generator, we set $h_u=512$, $k_u=[16, 16, 4, 4]$, $k_r=[3, 7, 11]$, and $D_r=[[[1,1], [3, 1], [5, 1]] \times 3]$ following \cite{kong2020hifi}. The networks were trained using the AdamW optimizer \cite{loshchilov2019d} with $\beta_1=0.8$, $\beta_2=0.99$, and weight decay $\lambda=0.01$. The initial learning rate was $2\times 10^{-4}$, decayed by a factor of $0.999$ every epoch. Our training was conducted on a single NVIDIA A800 GPU with a batch size of 16 for 500k steps.
\begin{figure}
  \centering
  \includegraphics[width=8.5cm]{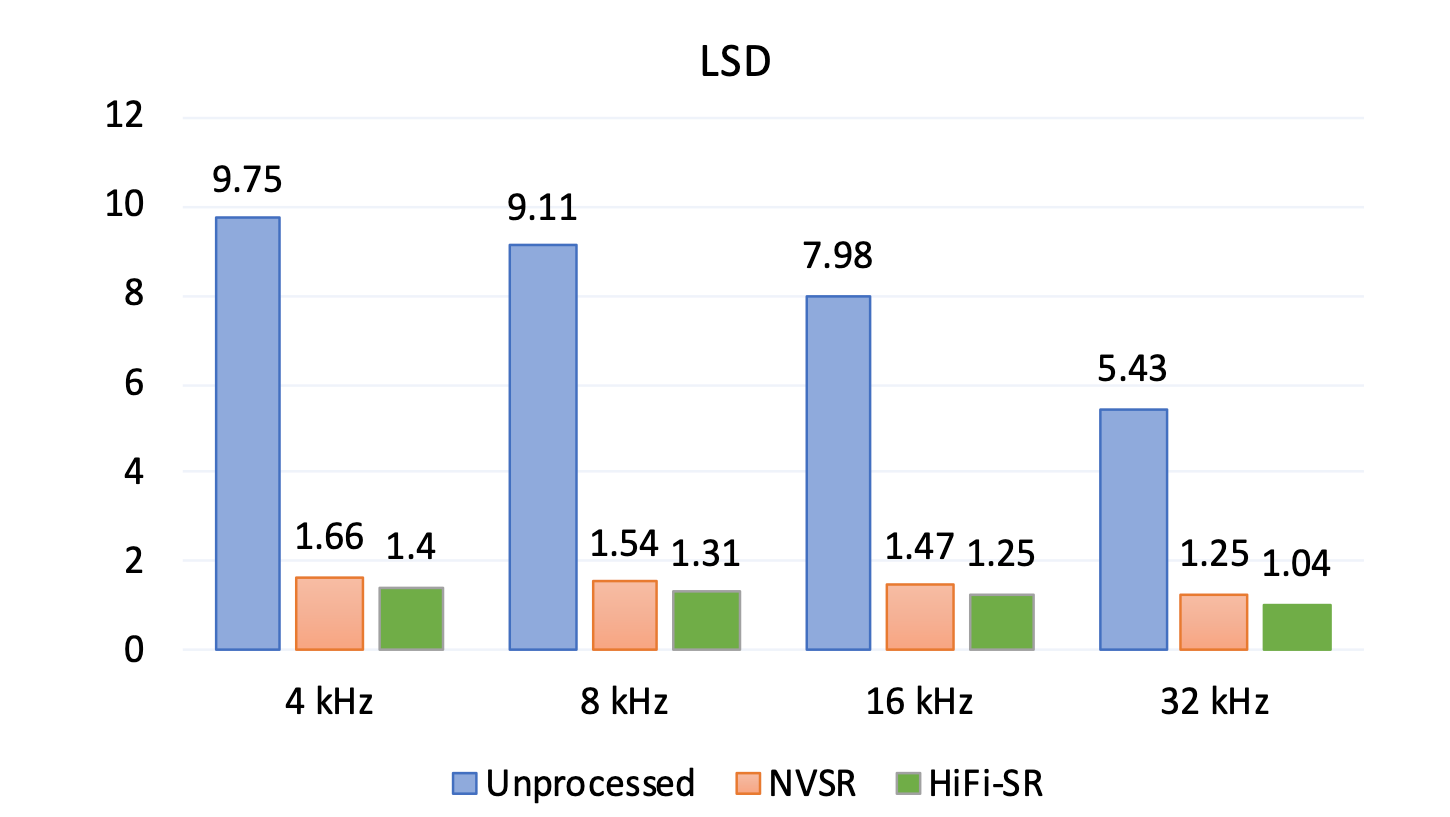}
  \vspace{-3mm}
  \caption{Comparison results of NVSR and HiFi-SR on VocalSet test set with 48 kHz target sampling rate and four input sampling rates. }
  \label{fig1}
\vspace{-3mm}
\end{figure} 
\begin{figure}
  \centering
  \includegraphics[width=8.0cm]{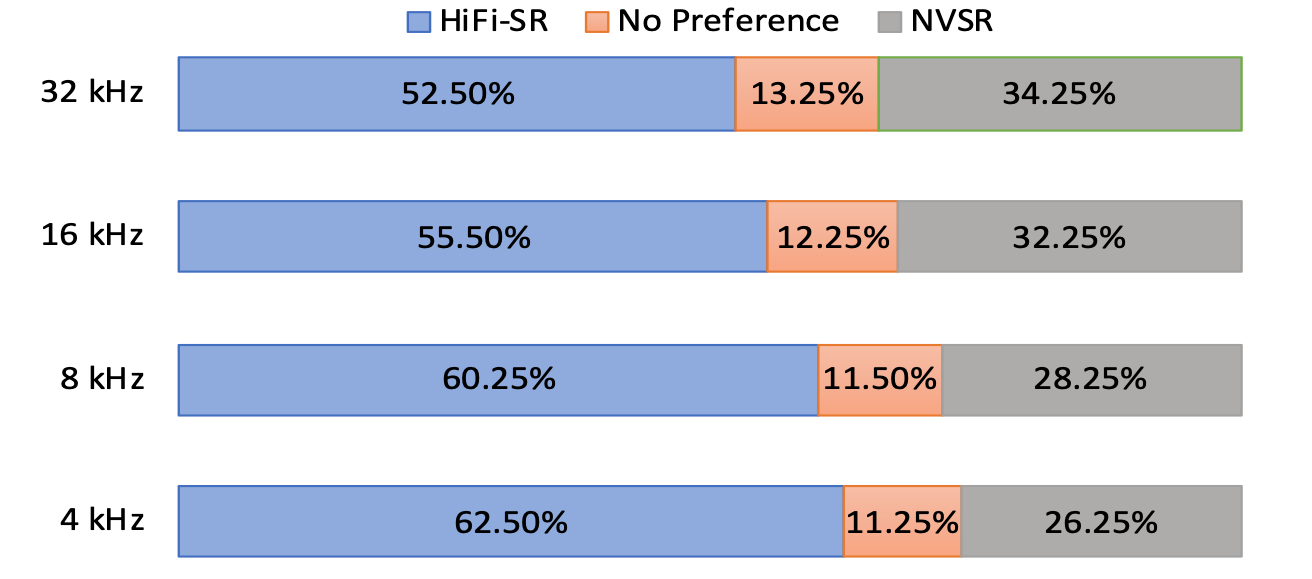}
  \vspace{-3mm}
  \caption{ABX subjective test results of NVSR and HiFi-SR on mixed EXPRESSO and VocalSet test set with 48 kHz target sampling rate and four input sampling rates. }
  \label{fig1}
\vspace{-3mm}
\end{figure} 
\subsection{Results and Discussion}
For objective evaluation, Table 1 compares the performance on the matched VCTK test set using the LSD metric. HiFi-SR achieves an average LSD of 0.82, outperforming all baseline models. The closest competitor is NVSR, with an average LSD of 0.85. We attribute this improvement to our proposed transformer-convolutional generator and adversarial training strategy. To further verify the effectiveness of our training strategies, we conducted ablation studies by removing MBD and the multi-scale mel-spectrogram loss $\mathcal{L}{m}(G)$. As shown in Table 1, without MBD, the average LSD slightly increases to 0.83, while removing $\mathcal{L}{m}(G)$ increases it to 0.84. On the EXPRESSO and VocalSet test sets, we compared HiFi-SR against the competitive NVSR model to assess generalization capabilities. The results, displayed in Figures 3 and 4, show that HiFi-SR outperforms NVSR by a larger margin on the out-of-domain test sets, demonstrating the superiority of our unified framework over NVSR’s separated-module approach.
%The results, displayed in

For subjective evaluation, the ABX test results are presented in Figure 5. We evaluated both the EXPRESSO and VocalSet test sets by randomly selecting 25 audio outputs from each set for both HiFi-SR and NVSR models, resulting in 50 audio pairs per sampling rate. Participants were asked to choose the audio output with better sound quality or indicate no preference. As shown in Figure 5, participants showed a higher preference for HiFi-SR audio outputs compared to NVSR, with HiFi-SR achieving a preference rate of over 52.50\% across all four input sampling rates. This demonstrates that our unified HiFi-SR model generalizes better than the NVSR model on out-of-domain test sets. We visualize the spectrograms of a processed sample from both NVSR and HiFi-SR in Figure 2. The output of HiFi-SR is noticeably closer to the ground truth.

\section{Conclusions}
In this paper, we presented HiFi-SR, a unified network developed to address the challenges of speech super-resolution, particularly in out-of-domain scenarios. By leveraging a transformer-convolutional generator and end-to-end adversarial training, HiFi-SR effectively handles both the prediction of latent representations and their conversion into time-domain waveforms, ensuring consistent and high-fidelity speech reconstruction. Our experimental results show that HiFi-SR outperforms existing speech SR methods, achieving superior performance in both objective metrics and ABX preference tests. The model's ability to generalize well to out-of-domain data further highlights the robustness of our approach.
% References should be produced using the bibtex program from suitable
% BiBTeX files (here: strings, refs, manuals). The IEEEbib.bst bibliography
% style file from IEEE produces unsorted bibliography list.
% -------------------------------------------------------------------------
\bibliographystyle{IEEEbib}
\bibliography{refs}

\end{document}